# Investigation of Surface State of Topological Kondo Insulator with Rashba Impurities

*Partha Goswami*

*D.B.College, University of Delhi, Kalkaji, New Delhi-110019, India*
*physicsgoswami@gmail.com*

**Abstract**

We study a generic topological Kondo insulator (TKI) system by performing a mean-field theoretic (MFT) calculation within the frame-work of slave-boson protocol. We assume infinite Hubbard-type interaction among the localized electrons. The difference between the bulk metallic and insulating phases of TKI is in the sign of nearest neighbor hopping of localized electrons: the hopping amplitude is positive for the metallic and negative for the insulating phase. The surface metallicity together with bulk insulation, however, requires very strong *f*-electron localization. Furthermore, we find that the exchange field, arising due to the presence of the magnetic impurities on the surface of the system, opens a gap at the gapless Dirac dispersion of the surface states. For the gapped surface state spectrum, we find the possibility of intra-band as well as inter-band unconventional plasmons. The paramountcy of the bulk metallicity, and, in the presence of the Rashba impurities, the TKI surface comprising of 'helical liquids' are the important outcomes of the present communication. The access to the gapless Dirac spectrum leads to spin-plasmons with the usual wave vector dependence $q^{1/2}$. The Rashba coupling does not impair the Kondo screening and does not affect the QCP for the bulk.

## 1. Introduction

The Periodic Anderson model (PAM) **[1-6, 25]** basically involves highly correlated electrons (localized magnetic moments) in one *f* orbital which are screened by weakly correlated electrons in a second *d* orbital. Although this model has been thoroughly examined for several decades, the model itself and its extensions **[5,6]** are still relevant for the theoretical condensed-matter physics. A captivating development in recent years is the discovery that Kondo insulators can develop topological order to form a topological Kondo insulator (TKI). The theoretical description of a Kondo insulator **[7-10, 16]** is usually based on PAM. In this paper we shall focus on a generic topological Kondo insulator with the simplest band structure. Upon including the effect of prototypical Rashba impurities, such as metals or metallic alloys, which bolster the Rashba spin-orbit interaction (RSOI) between *f* – electrons in the surface state Hamiltonian, we get access to the gapless Dirac spectrum and the spin-plasmons with the usual wave vector dependence $q^{1/2}$. The Kondo screening in the bulk remains unaffected by RSOI.

It may be mentioned that the first example of correlated topological Kondo insulator(TKI) **[1,7-10]** is $SmB_6$. It has a cubic crystal structure of the cesium-chloride type with a lattice constant of $a \approx 4.13$ Å. At high temperature, the material behaves like a metal, but when reducing the temperature below ∼ 50 K, it exhibits insulating behavior **[11]**. The hybridization gap (Δ) has been measured at ∼15 meV **[11,12]**. This constitutes typical behaviour for a Kondo insulator. However, $SmB_6$ shows some unusual properties in addition. It has been found **[13]** that the resistivity of $SmB_6$ increases like an insulator but saturates at temperatures below 5K upon decreasing the temperature. While non-magnetic impurities do not influence the saturation anyway, doping $SmB_6$ with magnetic impurities does **[14,15].** The unusual properties of this rare earth hexa-borides are ascribed to the interaction of the 5*d* and 4*f* electrons of rare earth element with the 2*p* conduction electrons of Boron. Unlike the well-known topological insulators, such as $Bi_2Se_3$ family of materials, in TKIs the strong correlation effects among the 4*f*-electrons are very important. It brings about the strong modification of the 4*f* -band width. It may be mentioned that in the $Bi_2Se_3$ family of materials, the band inversion happens between two bands both with the *p* character and similar band width. The situation in $SmB_6$ is quite different, where the band inversion happens between 5*d* and 4*f* bands with the band widths differing by several orders of magnitude, which leads to very unique low energy electronic structure.

To investigate the bulk model, the slave boson technique [1,17-20] is usually employed. The slave particle protocol is based on the assumption of spin-charge separation in the strongly correlated electron systems in Mott insulators. The surmise is that electrons can metamorphose into spinons and chargons. But to preserve the fermion statistics of the electrons, the spinon-chargon bound state must be fermionic, so the simplest way is to ascribe the fermion statistics to one of them: if the spinon is fermionic then the chargon should be bosonic (slave-boson), or if the chargon is fermionic then the spinon should be bosonic (slave-fermion). The two approaches are just two low-energy effective theories of the complete-fractionalization theory [21,22]. We use the slave-particle mean-field theory (MFT) [1,12-14] which allows the study of low-energy regime of a Kondo system with a quadratic single-particle Hamiltonian in the limit of the *f*-electron correlations being larger than all other energy scales in the problem. We obtain the self-consistent equations for MFT parameters minimizing the grand canonical potential of the system with relative to these parameters as in ref. [1]. The parameters enforce constraints on the pseudo-particles due to the infinite Coulomb repulsion and the need of the formation of singlet states between an itinerant electron and a localized fermion at each lattice site in order to have a Kondo insulator. The technique used by us to calculate various thermodynamic averages is the Green's function method as was used by Legner [1]. The theory allows to conveniently circumvent complications associated with formally, infinite repulsion between the *f*-electrons by "splitting" the physical *f*-electron into a product of a fermion and a slave boson, supplemented with a constraint to remove the double occupancy. The temperature range that slave-particle mean-field-theory is valid, limited to very low values (T → 0). Furthermore, MFT involves condensing the boson field and neglecting all of its dynamics. This effectively leads to a non-interacting model of an insulator, where the gap is proportional to the condensate and hybridization parameter, and makes it amenable to a topological analysis. Since the electron states being hybridized have the opposite parities, the resulting Kondo insulator is topological and contains its hallmark feature - the metallic surface states. Thus, the mean-field theory is reasonable, and works at the qualitative level. A fundamental quantity describing the magnetic response of the system is the spin susceptibility. The Kondo screening mechanisms are mostly characterized by the same response function. We wish to report the calculation of the bulk spin susceptibility of this system, identified by the dominance of the bulk metallic character, in a sequel to this work.

The paper is organized as follows: In section 2, we consider a model for a (topological) Kondo insulator on a simple cubic lattice with one spin-degenerate orbital per lattice site each for *d* and *f* electrons with Hubbard type interaction term (*U*) between the latter. We implement the slave-boson protocol to include the effect of infinite *U* into consideration at the mean-field theoretic (MFT) level. We obtain the grand canonical potential Ώ of the system in the slave-boson representation with infinite *U*. The mean-field parameters, such as slave-boson field, auxiliary chemical potential, and a Lagrange multiplier to enforce the prohibition of double occupancy were obtained by minimizing this grand canonical potential with respect to these parameters. In section 3, we investigate the surface state dispersion followed by the plasmonics of the surface states. Upon including the effect of Rashba impurities on the surface, we get access to the gapless Dirac spectrum and the spin-plasmons. The paper ends with a short discussion and concluding remarks at the end.

## 2. PERIODIC ANDERSON MODEL

The widely accepted model of a topological Kondo insulator (TKI) [7-10, 16] involves, alongside a strong spin-orbit coupling, hybridization between an odd-parity nearly localized band and an even-parity delocalized conduction band. In the case of $SmB_6$ with a cubic crystal structure, these bands correspond to 4*f* and 5*d* electrons, respectively. It is then imperative that we start with the periodic

Anderson model (PAM) where there are two different species of electrons, namely conduction electrons and localized electrons, often originating from *d* and *f* orbitals, respectively. We shall call them the conduction electrons and the valence electrons, respectively. The model **[1-6, 8, 25]** given below ignores the complicated multiplet structure of the *d* and *f* orbitals usually encountered in real TKIs such as $SmB_6$. This has no major implication as most topological properties of cubic Kondo insulators do not depend on the precise form of the hopping and hybridization matrix elements or, on the particular shape of the orbitals.

## 2.1 MODEL

We consider below a well-known model **[1-6,8]**for a (topological) Kondo insulator on a simple cubic lattice with one spin-degenerate orbital per lattice site each for *d* and *f* electrons. In momentum-space, we represent them by creation (annihilation) operators $d^\dagger_{k\zeta}$ ($d_{k\zeta}$) and $f^\dagger_{k\zeta}$ ($f_{k\zeta}$), respectively. Here, the index $\zeta$ ($=\uparrow,\downarrow$) represents the spin or pseudo-spin of the electrons, where the latter is relevant for the localized electrons with generally large SOC. The Hamiltonian consists of two parts, namely, the bare hopping of the individual orbitals plus the hybridization between *d* and *f* orbitals($\aleph$),and an onsite repulsion or *f* electrons($\aleph_{int}$). We have

$$\aleph = \sum_{k\zeta\,=\uparrow,\downarrow}(-\mu-\epsilon^d_k)d^\dagger_{k,\zeta}d_{k,\zeta} + \sum_{k,\zeta\,=\uparrow,\downarrow}(-\mu-\epsilon^f_k)f^\dagger_{k,\zeta}f_{k,\zeta}$$
$$+ \sum_{k,\zeta=\uparrow,\downarrow}\{\Gamma_{\zeta\,=\uparrow,\downarrow}(k)\,d^\dagger_{k\zeta}f_{k,\zeta}+\text{H.C.}\} \quad (1)$$

Where $\epsilon^d_k = [2t_{d1}\,c_1(k) + 4t_{d2}\,c_2(k) + 8t_{d3}\,c_3(k)]$ and $\epsilon^f_k = [-\epsilon_f + 2t_{f1}\,c_1(k) + 4t_{f2}\,c_2(k) + 8t_{f3}\,c_3(k)]$. The $(t_{d1},t_{f1})$, $(t_{d2},t_{f2})$, and $(t_{d3},t_{f3})$, respectively,are the *NN*, *NNN*, and *NNNN* hopping parameters. Also, $c_1(k) = (\cos k_x a + \cos k_y a + \cos k_z a)$, $c_2(k) = (\cos k_x a \cos k_y a + \cos k_y a \cos k_z a + \cos k_z a \cos k_x a)$, $c_3(k) = (\cos k_x a \cos k_y a \cos k_z a)$ with *a* as the lattice constant.The sums run over all values for the crystal momentum (*k*) and the index $\zeta=(\uparrow,\downarrow)$ for all (cubic) lattice sites. The dispersion of the *d* and *f* electrons is described by the first and the second term, respectively, and their hybridization or the (Dirac-type) spin-orbit interaction by the matrix $\Gamma_{\zeta=\uparrow,\downarrow}(k)$. Due to spin-orbit coupling the *f*-states are eigenstates of the total angular momentum *J*, and hence hybridize with conduction band states with the same symmetry. This gives rise to the momentum-dependence of the form-factor $\Gamma_{\zeta\,=\uparrow,\downarrow}(k)$. The Dirac-type spin–orbit interaction or hybridization term **[1, 21]** is given by $\Gamma_{\zeta\,=\uparrow,\downarrow}(k) = -2V(s(k)\cdot\zeta)$where $\zeta_\alpha$ are the Pauli matrices in physical spin space, and $s(k) = (\sin k_x a, \sin k_y a, \sin k_z a)$. The hybridization is the source of non-trivial topology of the emergent bands. Since the *f*- and *d*-states have different parities, for the hybridization we must have odd parity: $\Gamma_{\zeta\,=\uparrow,\downarrow}(-k) = -\Gamma_{\zeta\,=\uparrow,\downarrow}(k)$. The negative sign of $t_{f1}$ is necessary for the band inversion, which induces the topological state **[27]**.The system shows the insulating as well as the metallic phases. The difference between the metallic and insulating phase is the sign of $t_{f1}$: It is positive for the metallic and negative for the insulating phase. The bandwidth of the *f* electrons is much smaller than the bandwidth of the conduction electrons and we therefore assume that $|t_{f1}| \ll |t_{d1}|$ and similar relations hold for second- and third-neighbor hopping amplitudes. The hybridization is characterized by the parameter *V* for which we typically use $|V| < |t_{d1}|$. Throughout the whole paper, we choose $t_{d1}$ to be the unit of energy, $t_{d1}=1$. We note that the hybridization is an odd function of *k* in order to preserve time reversal symmetry(TRS), as it couples to the physical spin of the electron. The interaction term $\aleph_{int} = U\sum_i f^\dagger_{i\uparrow}f_{i\uparrow}f^\dagger_{i\downarrow}f_{i\downarrow}$ is the onsite repulsion of *f* electrons. Here we have assumed that the *f* electrons locally interact via a Hubbard-*U* repulsion while the *d* electrons are non-interacting. The total Hamiltonian (without the interaction term

U) yields the single–particle spectrum $\in_\alpha^{(\zeta)}(k) = -\frac{(\epsilon_k^d + \epsilon_k^f + \zeta M)}{2} + \alpha \sqrt{\frac{(\epsilon_k^d - \epsilon_k^f + \zeta M)^2}{4} + \in_h^2}$ where $\alpha = 1$ (−1) for upper band (lower band) $\zeta = \pm 1$ labels the eigenstates ( ↑, ↓ ) of $\zeta_z$, and $\in_h = -2V(s_x^2 + s_y^2 + s_z^2)^{1/2}$. The pictorial depiction of the spectrum is shown in Figure 1. It is clear from the figure that the bulk metallicity is accessible when $t_{f1}$ is positive, as the conduction bands in this case are half empty. On the other hand, when $t_{f1}$ is negative, the band gap exists between the valence and the conduction bands leading to the bulk insulation.

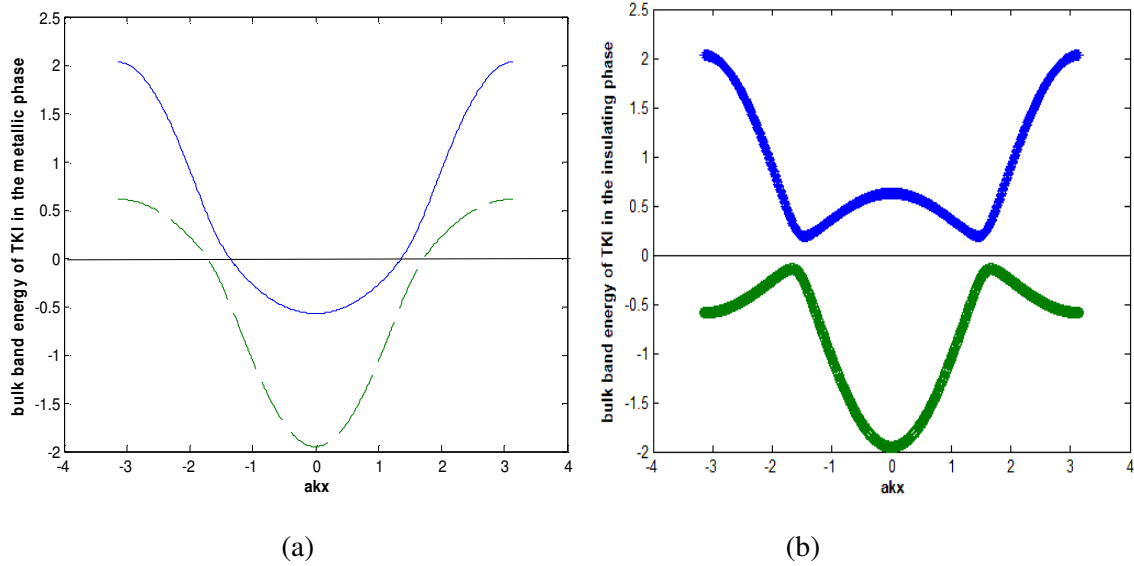

(a)          (b)

**Figure 1.** The two-band energy spectra of a bulk TKI. The parameter values are, $ak_y = \pi, t_{d1} = 1, t_{d2} = 0.01, t_{d3} = 0.001, t_{f2} = 0.01, t_{f3} = 0.001, \mu = 0.0, \epsilon_f = -0.02, V = 0.10$. In figure (a), $t_{f1} = 0.3$. The conduction band is partially empty which implies the bulk metallicity. In figure (b), $t_{f1} = -0.3$. In case this case, the Fermi level (Chemical potential represented by solid, horizontal line) is located in the hybridization gap which implies bulk insulation.

Our main aim in this paper is to capture physics associated with the inclusion of the Rashba impurities in the system, modeled by appropriate surface Hamiltonian derived from the bulk Hamiltonian ℵ + $Rashba\ coupling$ + ℵ$_{int}$. The interaction term ℵ$_{int}$ can be studied non-perturbatively by various methods, including dynamical mean-field theory, Gutzwiller-projected variational wave-functions, or slave-particle representations [1, 29,30]. In what follows we shall use the slave-particle protocol. The *f*-electron correlations, being larger than all other energy scales in the problem, effectively enforce a no-double-occupancy constraint on each site, and therefore contribute through virtual processes at low temperatures only. The original Hilbert space thus get projected onto a smaller subspace, where the double occupancy is excluded.

**2.2 SLAVE-BOSON PROTOCOL**

One utilizes the well-known slave-boson protocol [1,12-14] to do the projection onto the Hilbert space. In this protocol, the electron operator is expressed in terms of pseudo-fermions and slave-bosons. The operators $a^\dagger$ (heavy slave-boson creation operator) and $b^\dagger$ (light slave-boson creation operator), respectively, creates doubly-occupied and empty bosonic impurity states out of vacuum. The bosons created by $b^\dagger$ are supposed to carry the electron's charge. The single site fermionic occupation operator is denoted by $s^\dagger$. This corresponds to a fermionic operator and carry the electron's spin but no charge. For $U \rightarrow \infty$, the double occupancy is prohibited, and hence the operators $a^\dagger$ and $a$ will not be under consideration. The connection of the remaining auxiliary operators ($b, s$) to the physical $f$- electron operator is $f_\zeta^\dagger = s^\dagger{}_\zeta\, b$. In the slave-boson mean-field theory of the infinite-$U$ model ( which consists of replacing slave-boson field ($b$) at each lattice site by the modulus of its expecetation value – a $c$ number), the anti-commutation relation $\{f_\zeta, f_{\zeta'}^\dagger\} = \delta_{\zeta\zeta'}$, implies that $b^2 \{s_\zeta, s^\dagger{}_{\zeta'}\} = \delta_{\zeta\zeta'}$. To take care of the conservation of auxiliary particle number one needs to impose the restriction $\sum_\zeta s_\zeta^\dagger s_\zeta + b^\dagger b = 1$ or, $\sum_\zeta \langle s_\zeta^\dagger s_\zeta \rangle \cong 1 - b^2$ at a site. It is now straight-forward to write the thermal average of the TKI slave-boson mean-field Hamiltonian

$$\langle \aleph_{sb}(b,\lambda,\xi) \rangle = \sum_{k\zeta\,=\uparrow,\downarrow} (-\mu - \xi - \epsilon_{\boldsymbol{k}}^d)\, \langle d_{k,\zeta}^\dagger d_{k,\zeta} \rangle + \sum_{k,\zeta\,=\uparrow,\downarrow} (-\mu + \xi - b^2 \epsilon_{\boldsymbol{k}}^f + \lambda)\, \langle s_{k,\zeta}^\dagger s_{k,\zeta} \rangle$$

$$+ b \sum_{k,\zeta,\sigma\,=\uparrow,\downarrow} \{\Gamma_{\zeta\,=\uparrow,\downarrow}(\boldsymbol{k})\, \langle d_{k\zeta}^\dagger s_{k,\zeta} \rangle + H.C.\} + \lambda N_s (b^2 - 1) \quad (3)$$

where the additional terms, in comparison with (1), are $-\mu((N_d + N_s) - N) - \xi (N_d - N_s) + \lambda [\sum_{k,\sigma\,=\uparrow,\downarrow} \langle s_{k,\zeta}^\dagger s_{k,\zeta} \rangle + N_s (b^2 - 1)]$. The first term is the constraint which fixes the total number of particles $N$ ($N = N_d + N_s$), the second term enforces the fact that there are equal $d$ and $s$ fermions ($N_d = N_s$), and the third term describes the constraint on the pseudo-particles due to the infinite Coulomb repulsion. In order to have a Kondo insulator, formation of singlet states between $d$ and $s$ fermions is needed at each lattice site. This means that the number of $d$ and $s$ fermions are equal on average. Here $\lambda$ is a Lagrange multiplier, and $N_s$ is the number of lattice sites for $s$ electrons (similarly, $N_d$ corresponds to $d$-electrons). The dispersion of the $f$-electron is renormalized by $\lambda$ and its hopping amplitude by $b^2$. Moreover, the hybridization amplitude is also renormalized by the $c$-number $b$. The total parameters are slave-boson field $b$, auxiliary chemical potentials $\xi$ and $\mu$ ($\mu$ is a free parameter), and the Lagrange multiplier $\lambda$. Though these facts are explained clearly in ref. **[1]**, we have never-the-less found necessary to include them to make this paper self-contained. One obtains equations for the parameters ($b$, $\lambda$, $\xi$) minimizing the thermodynamic potential per unit volume $\Omega_{sb} = -(\beta V)^{-1} \ln Tr\, exp\, (-\beta \aleph_{sb}(b,\lambda,\xi))$ : $\partial \Omega_{sb}/\partial b = 0$, $\partial \Omega_{sb}/\partial \lambda = 0$, and $\partial \Omega_{sb}/\partial \xi = 0$. Here $\beta$ denotes the inverse of the product of temperature $T$ and Boltzmann constant $k_B$. The thermodynamic potential has been calculated by the method outlined in refs.**[23]** and **[24]**. These equations are :

$$2\lambda b = N_s^{-1} \sum_k \frac{\partial}{\partial b}[2V(s_x - i s_y)\langle d_{k\uparrow}^\dagger b s_{k,\downarrow}\rangle + 2V(s_x - i s_y)\langle b s^\dagger{}_{k,\uparrow} d_{k\downarrow}\rangle + H.C.]$$

$$+ N_s^{-1} \sum_{k,\zeta} \frac{\partial}{\partial b}[\epsilon_{\boldsymbol{k}}^f \langle b s_{k,\zeta}^\dagger b s_{k,\zeta}\rangle + \epsilon_{\boldsymbol{k}}^d \langle d_{k,\zeta}^\dagger d_{k,\zeta}\rangle] + N_s^{-1} \sum_{k,\zeta} \frac{\partial}{\partial b}[2V s_z \zeta \langle d_{k,\zeta}^\dagger b s_{k,\zeta}\rangle + H.C.], \quad (4)$$

$$(1 - b^2) = N_s^{-1} \sum_k \frac{\partial}{\partial \lambda}[2V(s_x - i s_y)\langle d_{k\uparrow}^\dagger b s_{k,\downarrow}\rangle + 2V(s_x - i s_y)\langle b s^\dagger{}_{k,\uparrow} d_{k\downarrow}\rangle + H.C.]$$

$$+ N_s^{-1} \sum_{k,\zeta} \frac{\partial}{\partial \lambda}[\epsilon_{\boldsymbol{k}}^f \langle b s_{k,\zeta}^\dagger b s_{k,\zeta}\rangle + \epsilon_{\boldsymbol{k}}^d \langle d_{k,\zeta}^\dagger d_{k,\zeta}\rangle] + N_s^{-1} \sum_{k,\zeta} \frac{\partial}{\partial \lambda}[2V s_z \zeta \langle d_{k,\zeta}^\dagger b s_{k,\zeta}\rangle + H.C.], \quad (5)$$

$$0 = N_s^{-1} \sum_k \frac{\partial}{\partial \xi}[2V(s_x - i s_y)\langle d_{k\uparrow}^\dagger b s_{k,\downarrow}\rangle + 2V(s_x - i s_y)\langle b s^\dagger{}_{k,\uparrow} d_{k\downarrow}\rangle + H.C.]$$

$$+ N_s^{-1} \sum_{k,\zeta} \frac{\partial}{\partial \xi}[\epsilon_{\boldsymbol{k}}^f \langle b s_{k,\zeta}^\dagger b s_{k,\zeta}\rangle + \epsilon_{\boldsymbol{k}}^d \langle d_{k,\zeta}^\dagger d_{k,\zeta}\rangle] + N_s^{-1} \sum_{k,\zeta} \frac{\partial}{\partial \xi}[2V s_z \zeta \langle d_{k,\zeta}^\dagger b s_{k,\zeta}\rangle + H.C.]. \quad (6)$$

$$\epsilon_k^d = [2t_{d1}\, c_1(k) + 4t_{d2}\, c_2(k) + 8t_{d3}\, c_3(k)]\,, \tag{7}$$

$$\epsilon_k^f = [-\epsilon_f + 2t_{f1}\, c_1(k) + 4t_{f2}\, c_2(k) + 8t_{f3}\, c_3(k)]\,, \tag{8}$$

The averages $\langle d_{k,\zeta}^\dagger d_{k,\zeta}\rangle, \langle bs_{k,\zeta}^\dagger bs_{k,\zeta}\rangle$, etc. have been calculated in appendix A below. Their expressions show that in the zero-temperature and the long-wavelength limits, the contribution of the averages $\langle d_{k\uparrow}^\dagger bs_{k,\downarrow}\rangle$ and $\langle bs^\dagger_{k,\uparrow} d_{k\downarrow}\rangle$, etc. to the derivatives in Eqs. (4), (5), and (6) are insignificant in comparison with those of $\sum_{k,\zeta}[\epsilon_k^f \langle bs_{k,\zeta}^\dagger bs_{k,\zeta}\rangle + \epsilon_k^d \langle d_{k,\zeta}^\dagger d_{k,\zeta}\rangle]$. This observation allows us to approximate the equations as

$$2\lambda b \approx N_s^{-1} \sum_{k,\zeta} \frac{\partial}{\partial b}[\epsilon_k^f \langle bs_{k,\zeta}^\dagger bs_{k,\zeta}\rangle + \epsilon_k^d \langle d_{k,\zeta}^\dagger d_{k,\zeta}\rangle]\,, \tag{9}$$

$$(1-b^2) \approx N_s^{-1}\sum_{k,\zeta}\frac{\partial}{\partial \lambda}[\epsilon_k^f \langle bs_{k,\zeta}^\dagger bs_{k,\zeta}\rangle + \epsilon_k^d \langle d_{k,\zeta}^\dagger d_{k,\zeta}\rangle]\,, \tag{10}$$

$$0 \approx N_s^{-1}\sum_{k,\zeta}\frac{\partial}{\partial \xi}[\epsilon_k^f \langle bs_{k,\zeta}^\dagger bs_{k,\zeta}\rangle + \epsilon_k^d \langle d_{k,\zeta}^\dagger d_{k,\zeta}\rangle] \tag{11}$$

in these limits. As we already have noted, for the conservation of auxiliary particle number, one needs to impose the restriction $\sum_\zeta \langle s_\zeta^\dagger s_\zeta\rangle \cong 1 - b^2$ at a site. On noting that $s_\zeta(r) = N_s^{-\frac{1}{2}}\sum_k e^{ik.r} s_{k,\zeta}$ where $N_s$ is the number of s-fermions, equivalently, one may also write this as $\int dr \sum_\zeta \langle bs_\zeta^\dagger(r) bs_\zeta(r)\rangle \cong N_s b^2(1 - b^2)$. This is the fourth equation with (9)-(11) as the first three and we have four unknowns, viz.( $b$, $\lambda$, $\xi$, $ak_F$) where $a \approx 4.13$ Å is the lattice constant and $k_F$ is the Fermi wave number. In view of the fact that the chemical potential is a free parameter and could be somewhere between the valence and the conduction bands, once again it is easy to see that for the low-lying states in the zero-temperature limit one may write the imposed restriction as $2 N_s^{-1}\sum_{k,\zeta} 1 = b^2 - b^4$. This is an equation for $b^2$ in terms of $(ak_F)$. After a little algebra, we find that whereas (9) and (10) together yield $\lambda = -6t_{f1} + 6b^2 t_{f1}$, Eq.(11) yields $\xi = -3t_{d1} + 3t_{f1}$. We estimate $(ak_F)$ in the following manner: Since the Fermi velocity $v_F^*$ of the low-lying states is known to be less than 0.2 eV-Å[25], taking the effective fermion mass ($m^* = \frac{\hbar k_F}{v_F^*}$) a hundred times that of an electron[26], we find that $(ak_F) \sim 0.01$. This is consistent with the long wavelength limit we have assumed. The two values of $b^2$ obtained from here are close to but less than one ($1^-$) and close to zero ($0^+$). The fact to remember is when $b$ is nonzero, the system is in Kondo state, and when $b$ vanishes, the system is in normal gas state. The admissible value of $b^2$ will be thus be $1^-$. With this a reasonable spectral gap, in the long wavelength limit, is obtained in the Kondo insulating state (see Figure 2(b)). We see in the figure that in order to have a gapped Kondo state it is required that the tunneling of the localized fermions must have an opposite sign compared with the conduction fermions. This is difficult to achieve, because the signs of the tunneling for the lowest Bloch bands are usually the same. However, it has been taken to be positive for the metallic phase and negative for the insulating phase for the following reason: For the metallic phase the leading term for $f$-electrons in the Hamiltonian, viz. $-2t_{f1}\, c_1(k)\, \langle bs_{k,\zeta}^\dagger bs_{k,\zeta}\rangle$, corresponds to a minimum at (0,0,0) (electron-like band) and in the insulating phase corresponds to a maximum (hole-like band). Furthermore, it must be mentioned that in their quantum simulation of the topological Kondo insulator in ultra-cold atoms, Zheng et al [27] have suggested a transformation leading to a staggered Kondo coupling – positive for the metallic phase and negative for the insulating phase. Abiding by this specification, we find that in the metallic phase $\xi$ is $-2.7$ and, in the insulating phase it is $-3.3$. All the unknown parameters now stand determined.

An investigation on Kondo system is incomplete without the possibility of the Kondo screening being explicitly shown. We, therefore, calculate the Kondo singlet density which may be defined as $K_{singlet}(k,b,\lambda,\mu,\xi) = [\langle d^\dagger_{k\uparrow} bs_{k,\downarrow}\rangle + \langle bs^\dagger_{k,\uparrow} d_{k\downarrow}\rangle]$. We obtain from the appendix A

$$K_{singlet}(k,\mu) = \frac{2V^2(s_x^2+s_y^2)}{\varepsilon_-(k,b,\lambda,\mu,\xi)}\left[\left(e^{\beta(\epsilon^{(-)}_-(k)-\mu)}+1\right)^{-1} - \left(e^{\beta(\epsilon^{(-)}_+(k)-\mu)}+1\right)^{-1}\right]$$

$$+ \frac{2V^2(s_x^2+s_y^2)}{\varepsilon_+(k,b,\lambda,\mu,\xi)}\left[\left(e^{\beta(\epsilon^{(+)}_-(k)-\mu)}+1\right)^{-1} - \left(e^{\beta(\epsilon^{(+)}_+(k)-\mu)}+1\right)^{-1}\right], \quad (12)$$

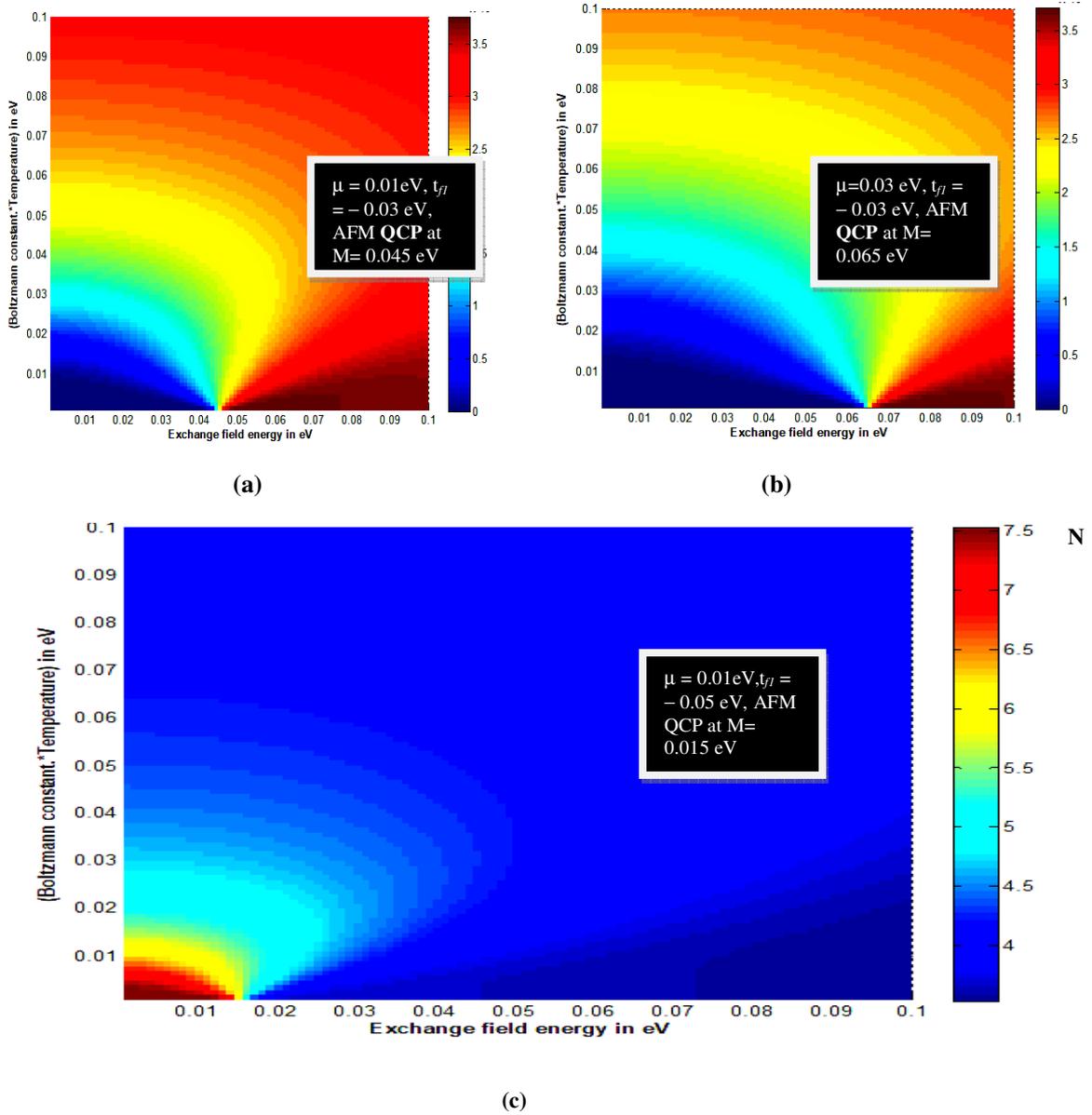

(a)

(b)

(c)

**Figure 2.** The contour plots of Kondo singlet term given by Eq.(12) as a function of exchange field energy ($M$) and (Boltzmann constant. Temperature)($kT$) in eV for **(a)** $\mu = 0.01$ eV, and $t_{f1} = -0.03$eV, **(b)** $\mu = 0.03$ eV, and $t_{f1} = -0.03$eV, and **(c)** $\mu = 0.01$ eV, and $t_{f1} = -0.05$eV at $ak_x = ak_y = ak_z = 1$. The anti-ferromagnetic quantum critical point (AFM QCP) is, respectively, at $M = M_{QCP} = 0.045$ eV, 0.065 eV, and 0.015 eV in (a), (b), and (c). Other parameters in the graphical

representations are $b = 0.95$, $t_{d1}= 0.5$ eV, $t_{d2}= 0.001$eV, $t_{f2}= 0.001$eV, $t_{d3}= 0.0001$eV, $t_{f3}= 0.0001$eV, $\varepsilon_f = -0.002$eV, and the hybridization parameter $V = 0.0001$ eV.

where $\varepsilon_+$, $\varepsilon_-$, $\in_\alpha^{(\zeta)}$, etc. are given by Eqs. (A7) to (A10). This average is the ultimate signature of the Kondo insulating state, where there is precisely one conduction electron paired with an impurity spin. The point we wish to make is that unless we have an anti-ferromagnetic exchange field in the bulk Hamiltonian this Kondo screening term will not be non-zero. Assuming $(V/t_{d1}) \ll 1$, we model the interaction between an impurity moment and the itinerant (conduction) electrons in the system with coupling term $-|J| \sum_m S_m \cdot s_m$, where $S_m$ is the $m$ th-site impurity spin, $s_m = \left(\frac{1}{2}\right) d^\dagger_{m\zeta} \zeta_z d_{m\zeta}$, $d_{m\zeta}$ is the fermion annihilation operator at site-$m$ and spin-state $\zeta$ $(=\uparrow,\downarrow)$ and $\zeta_z$ is the z-component of the Pauli matrices. We make the approximation of treating the impurity spins as classical vectors. The latter is valid for $|S| > 1$. Absorbing the magnitude of the impurity spin into the coupling constant $J$ $(M = |J||S|/t_{d1})$ it follows that the exchange field term, in the $(d_{k\uparrow}\ f_{k\uparrow}\ d_{k\downarrow}\ f_{k\downarrow})^T$ basis, appears as $\{\zeta_z \otimes M (\tau_0 + \tau_z)/2\}$, where $\tau_0$ and $\tau_z$, respectively, are the identity and the $z$-component of Pauli matrix for the pseudo-spin orbital indices. We thus obtain the dimensionless contribution $[M \sum_{k,\zeta} sgn(\zeta)\, d^\dagger_{k,\zeta} d_{k,\zeta}]$ to the momentum space Hamiltonian above. It is not difficult to see that the terms $\varepsilon_+$, $\varepsilon_-$, and $\in_\alpha^{(\zeta)}$ are to be redefined now in the following manner :

$$\varepsilon_-(k,b,\lambda,\xi) = \sqrt{\frac{(2\xi+\epsilon_k^d-b^2\epsilon_k^f+\lambda-M)^2}{4} + 4V^2 b^2(s_x^2 + s_y^2 + s_z^2)}, \tag{13}$$

$$\varepsilon_+(k,b,\lambda,\xi) = \sqrt{\frac{(2\xi+\epsilon_k^d-b^2\epsilon_k^f+\lambda+M)^2}{4} + 4V^2 b^2(s_x^2 + s_y^2 + s_z^2)}, \tag{14}$$

$$\in_\alpha^{(\zeta)}(k) = -\frac{(\epsilon_k^d+b^2\epsilon_k^f-\lambda+\zeta M)}{2} + \alpha\sqrt{\frac{(2\xi+\epsilon_k^d-b^2\epsilon_k^f+\lambda+\zeta M)^2}{4} + 4V^2 b^2(s_x^2 + s_y^2 + s_z^2)}, \tag{15}$$

$$\epsilon_k^d = [2t_{d1}\, c_1(k) + 4t_{d2}\, c_2(k) + 8t_{d3}\, c_3(k)],$$

$$\epsilon_k^f = [-\epsilon_f + 2t_{f1}\, c_1(k) + 4t_{f2}\, c_2(k) + 8t_{f3}\, c_3(k)], \tag{16}$$

Upon substituting these re-defined terms in Eq.(12) we obtain non-zero values of $K_{singlet}$. In figure 2 we have contour plotted $K_{singlet}$ as a function of the anti-ferromagnetic exchange field ($M$) and (Boltzmann constant. Temperature) ($kT$) in eV for **(a)** $\mu = 0.01$ eV, and $t_{f1}= -0.03$eV, **(b)** $\mu = 0.03$ eV, and $t_{f1}= -0.03$eV, and **(c)** $\mu = 0.01$ eV, and $t_{f1}= -0.05$eV at $ak_x = ak_y = ak_z = 1$. The anti-ferromagnetic quantum critical point (AFM QCP), respectively, is at $M = M_{QCP} = 0.045$ eV, 0.065 eV, and 0.015 eV in (a), (b), and (c)**.** This is typical Doniach-like phase diagram[32]**.** At T = 0K $M > M_{QCP}$ corresponds to the heavy-fermion liquid, while $M < M_{QCP}$ corresponds to anti-ferromagnetic liquid. From the plots we notice that the location of QCP depends on μ and $t_{f1}$. In fact, QCP increases with increase in μ and decreases with increase in $|t_{f1}|$.

In order to ascertain whether there is any effect of Rashba coupling (RC)($\lambda_R$) on the Kondo screening, we add the term $[2\lambda_R \sum_k [(k_y + ik_x) bs^\dagger_{k\uparrow} bs_{k,\downarrow} + H.C.]$ to the bulk Hamiltonian. We have assumed to have deposited particles with high Rashba spin-orbit (RSO) interactions $\lambda_R$, such as Au(111), on the bulk of material. We once again calculate the average $S = [2Vb(s_x - is_y)(\langle d^\dagger_{k\uparrow} bs_{k,\downarrow}\rangle + \langle bs^\dagger_{k,\uparrow} d_{k\downarrow}\rangle) + H.C.]$ corresponding to the Kondo singlet following the method outlined in the appendix A. We find that the RC does not impair the Kondo screening and does not affect the QCP for the bulk. Even in the surface state Hamiltonian in the next section, upon adding the Rashba coupling, one can show that the average similar to $S$ is non-zero. The stage is now set to investigate the surface state.

## 3. SURFACE STATE

### 3.1 Surface states

We choose a representation involving the states ($|c_{k,\uparrow,\pm}\rangle$, $|c_{k,\downarrow,\pm}\rangle$), where the operators $c_{k,\uparrow,\pm}$ and $c_{k,\downarrow,\pm}$, respectively, are the spin-up and spin-down annihilation operators corresponding to the upper and lower bands with spin-splitting ($\in^{(\uparrow)}{}_{\pm}(k)$ and $\in^{(\downarrow)}{}_{\pm}(k)$), as our basis. In the basis chosen above our bulk mean-field Hamiltonian matrix in slave-boson protocol will appear as

$$\begin{pmatrix} -\mu - \xi - \varepsilon_k^d & -2Vbs_z & 0 & -2Vb(s_x - is_y) \\ -2Vbs_z & -\mu + \xi - b^2\varepsilon_k^f + \lambda & -2Vb(s_x - is_y) & 0 \\ 0 & -2Vb(s_x + is_y) & -\mu - \xi - \varepsilon_k^d & 2Vbs_z \\ -2Vb(s_x + is_y) & 0 & 2Vbs_z & -\mu + \xi - b^2\varepsilon_k^f + \lambda \end{pmatrix} \quad (17)$$

The eigenvalues of this matrix are given by

$$\in_\alpha^{(\zeta)}(k) = -\frac{(\varepsilon_k^d + b^2\varepsilon_k^f - \lambda)}{2} + \alpha \sqrt{\frac{(2\xi + \varepsilon_k^d - b^2\varepsilon_k^f + \lambda)^2}{4} + 4V^2 b^2 (s_x^2 + s_y^2 + s_z^2)}, \quad (18)$$

$$\varepsilon_k^d = [2t_{d1} c_1(k) + 4t_{d2} c_2(k) + 8t_{d3} c_3(k)], \quad (19)$$

$$\varepsilon_k^f = [-\epsilon_f + 2t_{f1} c_1(k) + 4t_{f2} c_2(k) + 8t_{f3} c_3(k)], \quad (20)$$

. The energy eigenvectors corresponding to these eigenvalues $\left(\in_\alpha^{(\zeta)}(k)\right)$ are given by

$$\Psi_{k\alpha}^{(\zeta=-1)} = N_1^{-1/2} \begin{pmatrix} E_\alpha^{(\zeta=-1)}(k) \\ 2Vbs_z \\ 0 \\ 2Vb(s_x + is_y) \end{pmatrix},$$

$$\psi_{k\alpha}^{(\zeta=+1)} = N_2^{-1/2} \begin{pmatrix} 0 \\ -2Vb(s_x - is_y) \\ -E_\alpha^{(\zeta=1)}(k) \\ 2Vbs_z \end{pmatrix}. \quad (21)$$

where $(N_1, N_2) = \left[ E_\alpha^{(\zeta=\mp 1)^2}(\mathbf{k}) + 4V^2 b^2 \left( s^2_x + s^2_y + s^2_z \right) \right]^{\frac{1}{2}}$ correspond to the normalization terms and $E_\alpha^{(\zeta)}(\mathbf{k}) = \xi - b^2 \epsilon_k^f + \lambda - \epsilon_\alpha^{(\zeta)}(k)$. Our aim in this section is to examine the surface plasmons of the system under consideration. The plasmons are defined as longitudinal in-phase oscillation of all the carriers driven by the self-consistent electric field generated by the local variation in charge density on the surface. We require surface state single particle excitation spectrum for this purpose. To study the surface states, we first consider a thick slab limited in $z \in [-d/2, d/2]$ with open boundary conditions, where $d$ is the thickness of the slab in $z$ direction. In this case $k_z$ is not a good quantum number which should be replaced by $-i\partial_z$. The Hamiltonian $\aleph^{slab}(k_x, k_y)$ for the slab structure under consideration can be obtained from the Hamiltonian $\aleph^{bulk}(k_x, k_y, -i\partial_z)$ considering the ortho-normal function $| \varphi_n(z) \rangle = \psi_n(z)$ where $\psi_n(z) = \left( \frac{2}{d} \right)^{\frac{1}{2}} \left[ \sin\left\{ \frac{n\pi \left( z + \frac{d}{2} \right)}{d} \right\} \right]$ $(n = 0, 1, 2, 3, \ldots)$ ensuring $\psi_n \left( z = -\frac{d}{2}, \frac{d}{2} \right) = 0$. The matrix elements may be written as

$$\aleph_{mn}^{slab}(k_x, k_y) = \int_{-d/2}^{d/2} \langle \varphi_m(z) | \aleph_{mn}^{bulk}(k_x, k_y, -i\partial_z) \rangle | \varphi_n(z) \rangle dz. \tag{22}$$

The eigenstates of $\aleph_{mn}^{slab}(k_x, k_y)$ are the so called edge states. Equation (17) yields $\aleph_{mn}^{slab}$ as

$$\begin{pmatrix} \Gamma_1 & -\frac{16iVba}{3d} & 0 & 0 \\ \frac{16iVba}{3d} & \Gamma_2 & 0 & 0 \\ 0 & 0 & \Gamma_1 & \frac{96iVba}{7d} \\ 0 & 0 & -\frac{96iVba}{7d} & \Gamma_2 \end{pmatrix} \tag{23}$$

where

$$\Gamma_1(K) = -\mu - \xi - 6t_{d1} + t_{d1} K^2,$$

$$\Gamma_2(K) = -\mu + \xi + b^2 \varepsilon_f - 6b^2 t_{f1} + b^2 t_{f1} K^2 + \lambda. \tag{24}$$

Upon noting that $\lambda = -6t_{f1} + 6b^2 t_{f1}$ and $\xi \approx -3t_{d1}$, the eigenvalues of this matrix is given by

$$\epsilon_{\alpha, \, surface}^{(\zeta)}(K) = -\mu + \left( -3t_{d1} - 3t_{f1} + \frac{b^2}{2} \varepsilon_f \right) + \frac{t_{d1} + b^2 t_{f1}}{2}(Ka)^2 + \alpha$$

$$\times \sqrt{(CVb)^2 + \left( 3t_{f1} - \frac{b^2}{2} \varepsilon_f \right)^2 + v_F^*(\zeta)^2 (Ka)^2 + O((Ka)^4)} \tag{25}$$

where $v_F^*(\zeta) = \sqrt{(t_{d1} - b^2 t_{f1})\left( 3t_{f1} - \frac{b^2}{2} \varepsilon_f \right)}$ will be interpreted as the Fermi velocity. The constant $C = \left( \frac{16a}{3d} \right)$ and $\left( \frac{96a}{7d} \right)$, respectively, for the up and down quasi-particle spins. The eigenvectors corresponding to the eigenvalue $\epsilon_{\alpha, \, surface}^{(\zeta=+1)}(K)$ are

$$\tilde{V}^{(\zeta=+1)}_{k\alpha} = N_1^{-1/2} \begin{pmatrix} -\xi - b^2\varepsilon_f + 6b^2 t_{f1} - b^2 t_{f1} K^2 - \lambda + \epsilon^{(\zeta=+1)}_{\alpha,surface}(K) \\ 16Vbai/3d \\ 0 \\ 0 \end{pmatrix}, \qquad (26a)$$

and those corresponding to the eigenvalue $\epsilon^{(\zeta=-1)}_{\alpha,\,surface}(K)$ are

$$\tilde{V}^{(\zeta=-1)}_{k\alpha} = N_2^{-1/2} \begin{pmatrix} 0 \\ 0 \\ \xi + b^2\varepsilon_f - 6b^2 t_{f1} + b^2 t_{f1} K^2 + \lambda - \epsilon^{(\zeta=-1)}_{\alpha,surface}(K) \\ 96Vbai/7d \end{pmatrix}. \qquad (26b)$$

where $(N_1, N_2)$ correspond to the normalization terms. We notice that if the $f$-electrons somehow satisfy the condition $(3t_{f1} - \frac{b^2}{2}\varepsilon_f + \frac{(CVb)^2}{3t_{f1} - \frac{b^2}{2}\varepsilon_f})$ small compared to $(t_{d1} - b^2 t_{f1}) \times (aK)^2$, one obtains gapless surface state bands at (0,0) wavevector

$$\epsilon_{\alpha,\,surface}(K) \approx [-\mu + \left(-3t_{d1} - 3t_{f1} + \frac{b^2}{2}\varepsilon_f\right) + \alpha v_F^* \mid aK \mid + \frac{t_{d1} + b^2 t_{f1}}{2}(Ka)^2] \qquad (27)$$

for the insulating bulk. For this to happen, the severe restrictions are

$$\mid \varepsilon_f \mid > \frac{6\mid t_{f1}\mid}{b^2} \approx 6.5\mid t_{f1}\mid. \qquad (28a)$$

with $b^2 = 0.9890$ ( i.e. $\mid t_{f1}\mid$ should be less than $\mid \varepsilon_f \mid$ by one order of magnitude) and the high symmetry point (0,0) is unapproachable as

$$\mid aK \mid > \sqrt{\frac{3t_{f1} + v - \frac{b^2}{2}\varepsilon_f}{(t_{d1} - b^2 t_{f1})}}. \qquad (28b)$$

Here $v = \frac{(\frac{96a}{7d}Vb)^2}{3t_{f1} - \frac{b^2}{2}\varepsilon_f}$. The Fermi velocity $v_F^*$ is given by $\sqrt{(t_{d1} + b^2 \mid t_{f1}\mid)\left(-3\mid t_{f1}\mid + \frac{b^2}{2}\mid \varepsilon_f\mid\right)}$. With the parameter choice $ak_F = 0.01, t_{d1} = 500\,meV$, $\mid t_{f1}\mid = 5\,meV$, $t_{d2} = t_{d3} = t_{f2} = t_{f3} = 0$, $b^2 = 0.9890$, and $\mid \varepsilon_f \mid = 50$ meV, we find the estimated value $v_F^* \approx 10^4 m - s^{-1}$. The hybridization parameter $V$ cannot contribute here as we have chosen an ortho-normal function to obtain the Hamiltonian for the slab structure from the bulk Hamiltonian. It must be noted that Legner et al. [26] have found an expression of the Fermi velocity which with the same set of parameter values as above and $V = 100\,meV$ yields $v_F^*$ nearly the same value as obtained by us. The Kondo screening length $\xi\,(=\frac{\hbar v_F^*}{k_B T_{K,s}})$ for the surface states will be, therefore, be one order of magnitude higher than the lattice constant($a$) for the surface Kondo temperature $T_{K,s} \sim 25K$.

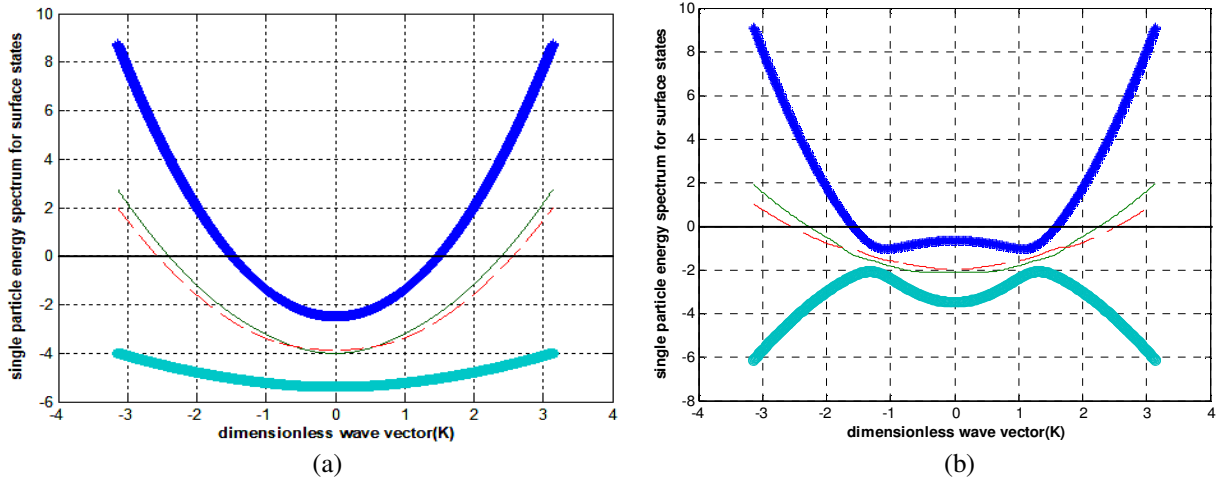

(a)                          (b)

**Figure 3.** The plots of single-particle excitation spectrum given by Eq.(25) versus dimensionless wave vector for chemical potential µ = 0 (unbroken horizontal straight line). The other parameters are b = 0.95, $t_{d1}$= 1, $\varepsilon_f$= -0.02, V = 0.10; $V_p$= $\left(\frac{Vab}{d}\right)$ = 0.04. The hopping integral (a) $t_{f1} = 0.31$ (metallic bulk) and $t_{f1} = -0.31$ (insulating bulk) .Since the conduction bands are partially empty, the surface state will be metallic in both (a) and (b).

We note that, with the exchange field introduced in the previous section, the eigenvalues of the matrix (23) is given by

$$\epsilon^{(\zeta)}_{\alpha,\ surface}(K) = -\mu + \left(-3t_{d1} - 3t_{f1} + \frac{b^2}{2}\varepsilon_f + \frac{\zeta M}{2}\right) + \frac{t_{d1}+b^2 t_{f1}}{2}(Ka)^2 + \alpha$$

$$\times \sqrt{(CVb)^2 + \frac{b^4}{4}\varepsilon_f^2 + \left(3t_{f1} + \frac{\zeta M}{2}\right)^2 - b^2\varepsilon_f\left(3t_{f1} + \frac{\zeta M}{2}\right) + v_F^*(\zeta)^2(Ka)^2 + O((Ka)^4)} \quad (29)$$

where $v_F^*(\zeta) = \sqrt{(t_{d1} - b^2 t_{f1})\left(3t_{f1} + \frac{\zeta M}{2} - \frac{b^2}{2}\varepsilon_f\right)}$ will now be interpreted as the Fermi velocity. It is thus demonstrated that the exchange field can be used to open an additional gap at the surface state dispersion.

We now consider the problem of the TKI surface plasmonics. The local variation in charge density in materials gives rise to an electric field. The quanta of in-phase longitudinal density oscillation of charge carriers at the surface of materials driven by this electric field are defined as the surface plasmons (SPs). Inside the bulk, SPs evanesce severely owing to the heavy energy loss. These collective density oscillations can be excited in the conventional metal surfaces. It is clear from the plot of Eq.(25) in Figure 3(b) that the surface could be metallic, even when the bulk is insulating, owing to the conduction bands being partially empty. This is, therefore, an appropriate case to investigate the possibility of the SPs. Equation (27) is expected to yield unconventional $q^{1/2}$ plasmons [36,37] obtainable under very stringent conditions given by (28). Since the calculation of surface metallicity is in the zero –temperature limit, a prediction of the previous works[7-10] that "when temperature is lowered a Kondo insulator may turn into a topological insulator with a metallic surface state" remains unverified. However, our finding of surface states with gapless Dirac dispersion albeit under very stringent conditions corroborates an important experimental finding of Xiang et al.[28]. They have found that, in the case of the prototypical TKI $SmB_6$, there is a broken rotating symmetry in the amplitude of the main de Haas–van Alphen oscillation branch consistent with Lifshitz-Kosevich theory confirming a 2D nature of the surface electronic state. The finding is similar to quantum oscillation experiments **[29,30]** for the conventional TIs to probe the surface Dirac fermions. The transport measurements **[31]** in the past have also demonstrated the insulating bulk and metallic surface separation. For potential applications toward scalable quantum information processing **[33]** this bulk and surface separation is especially important.

### 3.2 Plasmon frequency

The plasmons are defined as longitudinal in-phase oscillation of all the carriers driven by the self-consistent electric field generated by the local variation in induced charge density $\rho(r,\omega)$. In a linear-response approximation, we have $\rho(r,\omega) = e^2 \int d^2r' \chi(r,r',\omega) \Phi(r',\omega)$ where $\Phi$ is induced local potential and $\chi$ is the fermion response function or the dynamic polarization. This is a quantity of interest for many physical properties, since it determines e.g. the plasmon and phonon spectra. Assuming plasmon oscillation for the 3D system under consideration entirely a surface phenomenon, in the random phase approximation (RPA) **[34,35]**, we write the dynamical polarization function $\chi(a\boldsymbol{q},\omega)$ in the momentum space, as

$$\chi(a\boldsymbol{q},\omega) = \sum_{K,\zeta,\zeta',\alpha,\alpha'} |\langle \Psi_{\zeta,\alpha}(a(\boldsymbol{K}-\boldsymbol{q}))| \Psi_{\zeta',\alpha'}(a\boldsymbol{K})\rangle|^2 \left[\frac{n_{\zeta,\alpha}(a\boldsymbol{K}-a\boldsymbol{q}) - n_{\zeta',\alpha'}(a\boldsymbol{K})}{\{\hbar\omega + \epsilon_{\zeta,\alpha,surface}(a\boldsymbol{K}-a\boldsymbol{q}) - \epsilon_{\zeta',\alpha',surface}(a\boldsymbol{K}) + i\eta\}}\right]. \tag{30}$$

The symbol $\epsilon_{\zeta,\alpha,surface}(a\boldsymbol{K})$ stands for the surface state single-particle excitation spectrum given by (25) and $|\langle \Psi_{\zeta,\alpha}(a(\boldsymbol{K}-\boldsymbol{q}))| \Psi_{\zeta',\alpha'}(a\boldsymbol{K})\rangle|^2$ for the band-overlap of wave functions. This part of the paper leans on the previous investigations **[36,37]**. Since we are presently interested on intra-band plasmons only, we write down the explicit expression for the intra-band overlap. In view of (26), this is given by $F_{\alpha,\alpha,\zeta,\zeta'}(K,q) = \left(\frac{1}{2}\right) [1 + \zeta\zeta' \cos\theta_{\alpha,\alpha,K,q}]$, where

$$\cos\theta_{\alpha,\alpha,K,q,} = \left[\frac{(aq)\left(\frac{\partial \epsilon_{\alpha,surface}(K)}{\partial K} - \frac{\partial \Gamma_2(K)}{\partial K}\right)(\epsilon_{\alpha,surface}(K) - \Gamma_2(K))}{\left\{\left(\epsilon_{\alpha,surface}(K) - \Gamma_2(K)\right)^2 + (CVb)^2\right\}}\right]. \tag{31}$$

in the long wavelength limit and $\zeta\zeta' = +1$. The Fermi function with the band index α is given by $n_{\zeta,\alpha}(a\boldsymbol{K}) = 1/[exp(\beta(\epsilon_{\zeta,\alpha,surface}(a\boldsymbol{K})) - \beta\mu) + 1]$. Upon using the Sokhotski-Plemelj identity $(x \pm i\eta)^{-1} = P(x^{-1}) \mp i\pi\delta(x)$ with $P$ as the principal part, the real-part of the polarization function appears as

$$\chi_1(a\boldsymbol{q},\omega) = P \sum_{K,\alpha,\alpha',\zeta,\zeta'} \left[\frac{(n_{\zeta,\alpha}(a\boldsymbol{K}-a\boldsymbol{q}) - n_{\zeta',\alpha'}(a\boldsymbol{K}))F_{\alpha,\alpha',\zeta,\zeta'}(K,q)}{\{\hbar\omega + \epsilon_{\zeta,\alpha,surface}(a\boldsymbol{K}-a\boldsymbol{q}) - \epsilon_{\zeta',\alpha',surface}(a\boldsymbol{K})\}}\right]. \tag{32}$$

The imaginary part is given by

$$\chi_2(a\delta\boldsymbol{q},\omega) = -\pi \sum_{K,\alpha,\alpha',\zeta,\zeta'} (n_{\zeta,\alpha}(a\boldsymbol{K}-a\boldsymbol{q}) - n_{\zeta',\alpha'}(a\boldsymbol{K}))F_{\alpha,\alpha',\zeta,\zeta'}(K,q) \times \delta(\hbar\omega + \epsilon_{\zeta,\alpha,surf}(a\boldsymbol{K}-a\boldsymbol{q}) - \epsilon_{\zeta',\alpha',surf}(a\boldsymbol{K})). \tag{33}$$

Since we have chosen $t_{d1}$ to be the unit of energy before, the quantity $\hbar\omega$ is also in the same unit. In the long-wavelength limit, the band structure in Eq.(22) yields

$$\epsilon_{\zeta,\alpha,surface}(a\mathbf{K}-a\mathbf{q}) - \epsilon_{\zeta,\alpha,surface}(a\mathbf{K}) \approx \left\{\frac{a^2(q^2 - 2\mathbf{q}.\mathbf{K})}{2\lambda_{\zeta,\alpha}}\right\}. \tag{33}$$

where

$$\lambda_{\zeta,\alpha} \approx \frac{\frac{2}{t_{d1}}}{\left\{1+\frac{b^2 t_{f1}}{t_{d1}} + \frac{\alpha v_F^*(\zeta)^2}{2t_{d1}\left[(CVb)^2+\left(\frac{b^2}{2}\varepsilon_f - 3t_{f1}\right)^2\right]^{\frac{1}{2}}}\right\}} \approx \frac{2}{t_{d1}}\left(1 - \frac{b^2 t_{f1}}{t_{d1}}\right). \tag{34}$$

Within the random phase approximation (RPA), the plasmon dispersion is obtained by finding zeros of the dynamical dielectric function, which is expressed in terms of Coulomb's potential as $e_{\zeta,\alpha}(a|\mathbf{q}|,\omega') = 1 - V(q)\chi_{\zeta,\alpha}(a|\mathbf{q}|,\omega')$ where $\omega' = \omega - i\gamma$, $\gamma$ is the decay rate of plasmons, the expression $V(q)$ is the Fourier transform of the Coulomb potential in two dimensions. For weak damping, the equation $Re\ e(\omega, a|\mathbf{q}|) = 0$ yields the plasmon frequency $\hbar\omega_{pl}$. In the intra-band case, from Eq.(32), in the high frequency limit we obtain

$$e_{\zeta,\alpha}(a|\mathbf{q}|,\omega) = 1 - V_0 \sum_{\mathbf{K},\alpha,\zeta}\left[\frac{(\hbar\omega)^{-1}\frac{\partial n_{\zeta,\alpha}}{\partial\mu}\frac{\partial \epsilon_{\zeta,\alpha,surface}(a\mathbf{K})}{\partial(a\mathbf{K})}F_{\alpha,\zeta}(\mathbf{K},q)}{\left\{1-\left\{\frac{(a\mathbf{q}.a\mathbf{k})}{\hbar\omega\ \lambda_{\zeta,\alpha}}\right\}\right\}}\right]. \tag{35}$$

where $V_0 = \left(\frac{e^2}{2\varepsilon_0 \varepsilon_r}\right)$, $\varepsilon_0$ is the vacuum permittivity, and $\varepsilon_r$ is the relative permittivity of the surrounding medium.. The denominator of the summand in (35) involves a scalar product. We make use of the standard integral

$$\int_0^{2\pi} \frac{d\varphi}{\{a - x\cos\varphi\}} = 2\pi(2\theta(a)-1)\left(|a|^2 - x^2\right)^{-1/2}, \text{ for } |a| > x, \tag{36}$$

and zero, for $|a| \leq x$. Here $\theta(a)$ is the Heaviside step function. We can write $\int_0^{2\pi} \frac{d\varphi}{\left\{a-\left\{\frac{a|\mathbf{q}|a|\mathbf{k}|}{\hbar\omega\ \lambda_{\zeta,\alpha}}\right\}\cos\varphi\right\}}$

$=2\pi\left\{1-\left(\frac{a|\mathbf{q}|a|\mathbf{k}|}{\hbar\omega\ \lambda_{\zeta,\alpha}}\right)^2\right\}.^{-1/2}$ This integral allow us to write Eq.(35) as

$$e_{\zeta,\alpha}(a|\mathbf{q}|,\omega) = 1 - 2\pi V_0 \sum_{\mathbf{K},\alpha,\zeta}\left[\frac{(\hbar\omega)^{-1}\frac{\partial n_{\zeta,\alpha}}{\partial\mu}\frac{\partial \epsilon_{\zeta,\alpha,surface}(a\mathbf{K})}{\partial(a\mathbf{K})}F_{\alpha,\zeta}(\mathbf{K},q)}{\left\{1-\left(\frac{a|\mathbf{q}|a|\mathbf{k}|}{\hbar\omega\ \lambda_{\zeta,\alpha}}\right)^2\right\}^{1/2}}\right]. \tag{37}$$

The denominator of the summand in (37) can be expanded using the Binomial theorem. We obtain then the following implicit equation as the Plasmon dispersion for the system:

$$(\hbar\omega)^3 - (\hbar\omega)^2\left\{\pi V_0 \int d\mathbf{K} \sum_{\alpha,\zeta} \frac{\partial n_{\zeta,\alpha}}{\partial\mu}\frac{\partial \epsilon_{\zeta,\alpha,surface}(a\mathbf{K})}{\partial(a\mathbf{K})}\ (1+(aq)\mathbb{F}_K)\right\}$$

$$= \frac{1}{2}\pi V_0 (aq)^2 \int d\mathbf{K} \sum_{\alpha,\zeta} \frac{\partial n_{\zeta,\alpha}}{\partial \mu} \frac{\partial \epsilon_{\zeta,\alpha,surface}(aK)}{\partial (aK)} \left(\frac{a|\mathbf{K}|}{\lambda_{\zeta,\alpha}}\right)^2 (1 + (aq)\mathbb{F}_K), \qquad (38)$$

where $d\mathbf{K} = \left(\frac{d^2(a\mathbf{K})}{(2\pi)^2}\right)$ and

$$\mathbb{F}_K = \left[\frac{\left(\frac{\partial \epsilon_{\alpha,surface}(K)}{\partial K} - \frac{\partial \Gamma_2(K)}{\partial K}\right)\left(\epsilon_{\alpha,surface}(K) - \Gamma_2(K)\right)}{\left\{\left(\epsilon_{\alpha,surface}(K) - \Gamma_2(K)\right)^2 + (CVb)^2\right\}}\right]. \qquad (39)$$

It may be noted that we have put $\zeta\zeta' = +1$ in $F_{\alpha,\alpha,\zeta,\zeta'}(K,q) = \left(\frac{1}{2}\right)\left[1 + \zeta\zeta' \cos\theta_{\alpha,\alpha,K,q}\right]$ and wrote it as $(1 + (aq)\mathbb{F}_K)$ in Eq.(38), for we are considering the intra-band case presently. The integrals in (38) are trivial as $\frac{\partial n_{\zeta\alpha}(aK)}{\partial \mu}$ could be replaced by a delta function at T = 0K. Equation (38) may further be written as

$$(\hbar\omega)^3 - (\hbar\omega)^2 \pi V_0 (I_1 + (aq)I_2) - \frac{1}{2}\pi V_0 (aq)^2 (I_3 + (aq)I_4) = 0, \qquad (40)$$

where

$$I_1 = \left\{\int d\mathbf{K} \sum_{\alpha,\zeta} \frac{\partial n_{\zeta,\alpha}}{\partial \mu} \frac{\partial \epsilon_{\zeta,\alpha,surface}(aK)}{\partial (aK)}\right\}, \quad I_2 = \left\{\int d\mathbf{K} \sum_{\alpha,\zeta} \frac{\partial n_{\zeta,\alpha}}{\partial \mu} \frac{\partial \epsilon_{\zeta,\alpha,surface}(aK)}{\partial (aK)} \mathbb{F}_K\right\},$$

$$I_3 = \int d\mathbf{K} \sum_{\alpha,\zeta} \frac{\partial n_{\zeta,\alpha}}{\partial \mu} \frac{\partial \epsilon_{\zeta,\alpha,surface}(aK)}{\partial (aK)} \left(\frac{a|\mathbf{K}|}{\lambda_{\zeta,\alpha}}\right)^2, \quad I_4 = \int d\mathbf{K} \sum_{\alpha,\zeta} \frac{\partial n_{\zeta,\alpha}}{\partial \mu} \frac{\partial \epsilon_{\zeta,\alpha,surface}(aK)}{\partial (aK)} \left(\frac{a|\mathbf{K}|}{\lambda_{\zeta,\alpha}}\right)^2 \mathbb{F}_K.$$
$$(41)$$

Upon solving (40), one finds unconventional dispersion relation, viz.

$$\hbar\omega \approx \frac{3 - 2^{\frac{1}{3}}}{9} \pi V_0 (I_1 + (aq)I_2) \qquad (42)$$

in the long wavelength limit. To the leading order, $I_1$ and $I_2$ are positive as $I_n \sim \frac{t_{d1}(aK_F)^{n+2}}{n+2}$. It is to be noted that, though the relation is linear, the group velocity ($v_g$) and the phase velocity ($v_p$) are not equal. The former, given by $v_g = (0.6075 \, V_0 I_2)$ is several order of magnitude smaller than the speed of light whereas the latter is positive and much larger than unity. Thus, considering the intra-band transitions only, we have found possibility of only one collective mode exhibiting linear dispersion which could be triggered into an excited state with very low levels of energy input – less than 1 electron-volt.. It corresponds to charge plasmons and not spin plasmons as its origin does not lie in the gapless Dirac spectrum. The linear behaviour of the dispersion implies that signals can be transmitted undistorted along the surface. The finding has significant importance in extremely low loss communications.

As regards the inter-band plasmons corresponding to a pair of spin-split bands, we shall have $\zeta\zeta' = -1$. The band-overlap of wave functions $F_{\alpha,\alpha,\zeta,\zeta'}(K,q) = \left(\frac{1}{2}\right)\left[1 + \zeta\zeta' \cos\theta_{\alpha,\alpha,K,q}\right]$ may be written as $\left(\frac{1}{2}\right)(1 - (aq)\mathbb{F}_K)$. Furthermore, the quantity $\varepsilon_r$ is the relative permittivity of the surrounding medium and it is positive (negative) for a meta-material(negative dielectric constant medium). We assume the surrounding medium to be a meta-material. Thus, we may write $V_0 = -|V_0|$. Equation (42) in this case may be written as $\hbar\omega \approx \frac{3-2^{\frac{1}{3}}}{9}\pi V_0(-I_1+(aq)I_2)$. As in the case of frequency dispersion in groups of gravity waves on the surface of deep water, we encounter the exotic possibility of the group velocity being positive, while the phase velocity is negative. Our analysis demonstrates that the same plasmonic system can support both type of solutions, viz. $(v_g, v_p) > 0$ and $(v_g > 0, v_p < 0)$, depending on parameters.

### 3.3 Surface spectrum with Rashba coupling

In the previous sub-section we have obtained the gapless Dirac spectrum (27) under the severe restrictions (28) and (29). If this is true, we expect to have the well-known $q^{1/2}$ spin-plasmons **[36,37]**. Since the restrictions are ironclad, we look for other means to obtain gapless spectrum.. None other than the spin-orbit coupling is most suitable for the purpose. We, therefore, propose the consideration of an additional( Rashba) term $[2b^2\lambda_R(k_y + ik_x) + H.C]$ in the surface state Hamiltonian. It is imperative to assume that there must be deposition of particles with considerably high Rashba spin-orbit (RSO) interaction strength $\lambda_R$, such as Au(111), on the surface of material. The effect of this intrinsic coupling is to be included in the band -structure only. To start with, we consider Eq.(23) for $\aleph_{mn}^{slab}$ and add Rashba term to it. We obtain $\aleph_{mn}^{modified}$ as

$$\begin{pmatrix} \Gamma_1 & -\frac{16iVba}{3d} & 0 & 0 \\ \frac{16iVba}{3d} & \Gamma_2 & 0 & 2b^2\lambda_R(k_y + ik_x) \\ 0 & 0 & \Gamma_1 & \frac{96iVba}{7d} \\ 0 & 2b^2\lambda_R(k_y - ik_x) & -\frac{96iVba}{7d} & \Gamma_2 \end{pmatrix} \quad (43)$$

where

$$\Gamma_1(K) = -\mu - \xi - 6t_{d1} + t_{d1}K^2,$$

$$\Gamma_2(K) = -\mu + \xi + b^2\varepsilon_f - 6b^2 t_{f1} + b^2 t_{f1}K^2 + \lambda. \quad (44)$$

Upon noting that $\lambda = -6t_{f1} + 6b^2 t_{f1}$ and $\xi \approx -3t_{d1}$, the eigenvalues ε of this matrix is given by the quartic

$$(\varepsilon^2 - (\Gamma_1 + \Gamma_2)\varepsilon + \Gamma_1\Gamma_2 - \left(\frac{96Vba}{7d}\right)^2) \times (\varepsilon^2 - (\Gamma_1 + \Gamma_2)\varepsilon + \Gamma_1\Gamma_2 - \left(\frac{16Vba}{3d}\right)^2)$$

$$-(2b^2\lambda_R)^2(\varepsilon - \Gamma_1)^2 K^2 = 0, \quad (45)$$

where $K = \sqrt{(k_y^2 + k_x^2)}$. After lengthy algebra, Eq.(45) yields the following roots in view of the Ferrari's solution of a quartic equation:

$$\varepsilon_{s,\sigma}(K) = \sigma\sqrt{\frac{z_0(K)}{2} + \frac{(\Gamma_1(K)+\Gamma_2(K))}{2} + s\left(b_0(K) - \left(\frac{z_0(K)}{2}\right) + \sigma\, c_0(K)\sqrt{\frac{2}{z_0(K)}}\right)^{\frac{1}{2}}}, \quad (46)$$

where $\sigma = \pm 1$ is the spin-index and $s = \pm 1$ is the band-index. The other functions appearing in (46) are defined below:

$$z_0(K) = \frac{2b_0(K)}{3} + \left(\frac{1}{2}\Delta^{\frac{1}{2}}(K) - A_0(K)\right)^{\frac{1}{3}} - \left(\frac{1}{2}\Delta^{\frac{1}{2}}(K) + A_0(K)\right)^{\frac{1}{3}}, \quad (47)$$

$$A_0(K) = \left(\frac{b_0^3(K)}{27} - \frac{b_0(K)d_0(K)}{3} - c_0^2(K)\right), \; b_0(K) = \frac{3B^2(K)-8C(K)}{16}, \; c_0(K) = \frac{-B^3(K)+4B(K)C(K)-8D}{32}, \quad (48)$$

$$d_0(K) = \frac{-3B^4(K)+256E(K)-64B(K)D(K)+16B^2(K)C(K)}{256}, \quad (49)$$

$$\Delta(K) = \left(\frac{8}{729}b_0^6 + \frac{16d_0^2 b_0^2}{27} + 4c_0^4 - \frac{4d_0 b_0^4}{81} - \frac{8c_0^2 b_0^3}{27} + \frac{8c_0^2 b_0 d_0}{3} + \frac{4}{27}d_0^3\right), \quad (50)$$

$$B(K) = -2\big(\Gamma_1(K) + \Gamma_2(K)\big), \quad (51)$$

$$C(K) = \left[\big(\Gamma_1(K) + \Gamma_2(K)\big)^2 + 2\,\Gamma_1(K)\,\Gamma_2(K) - \left(\frac{96V_p}{7}\right)^2 - \left(\frac{16V_p}{3}\right)^2 - (2b^2\lambda_R)^2 K^2\right], \quad (52)$$

$$D(K) = -\big(\Gamma_1(K) + \Gamma_2(K)\big)\left(2\,\Gamma_1(K)\,\Gamma_2(K) - \left(\frac{96V_p}{7}\right)^2 - \left(\frac{16V_p}{3}\right)^2\right) + 2(2b^2\lambda_R)^2 \Gamma_1(K) K^2, \quad (53)$$

$$E(K) = \left(\Gamma_1(K)\,\Gamma_2(K) - \left(\frac{96V_p}{7}\right)^2\right)\left(\Gamma_1(K)\,\Gamma_2(K) - \left(\frac{16V_p}{3}\right)^2\right) - (2b^2\lambda_R)^2 \Gamma_1^2(K) K^2. \quad (54)$$

Moving over to Eq.(46), we notice that the first term $\sqrt{(z_0/2)}$ acts as an in-plane Zeeman term $g_b\mu_B B$ (where $B$ is the magnetic field, and $g_b$ is the Lande g-factor, and $\mu_B$ is the Bohr magneton). The pseudo-Zeeman term of the spectrum (46) comes into being due the presence of the term $(2b^2\lambda_R)^2(\varepsilon - \Gamma_1)^2 K^2$ in (45). Without this term, the spectrum reduces to a bi-quadratic (with no Zeeman term) rather than a quartic. Thus, the role of Rashba SOI as the polarization-usherer could be easily understood.

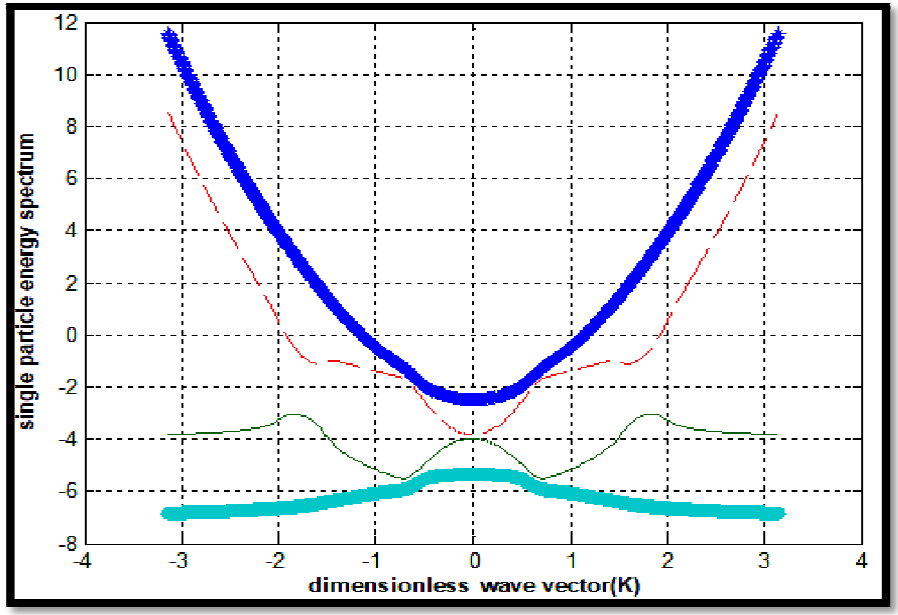

(a)

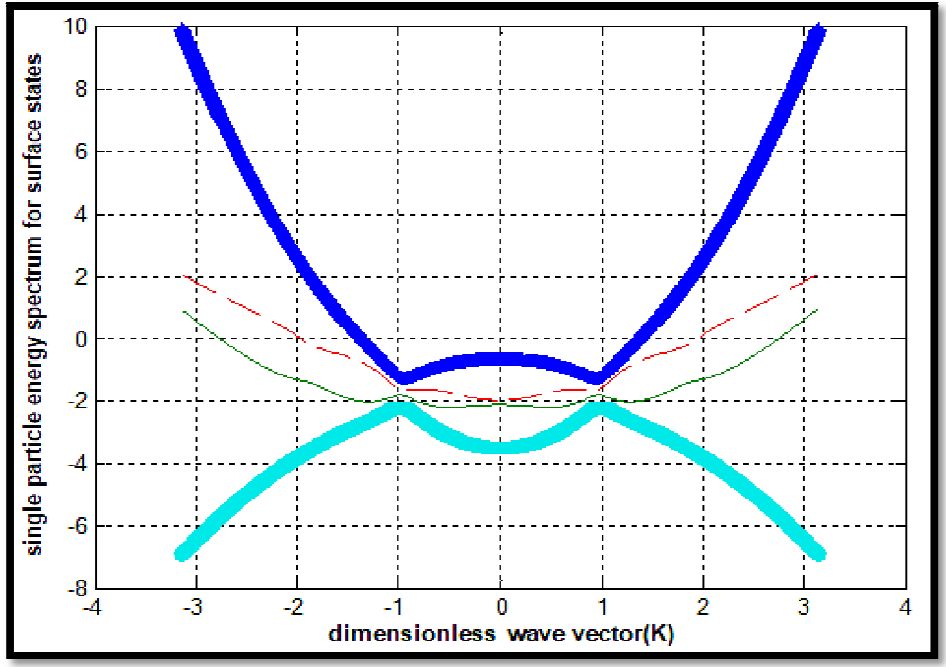

**(b)**

**Figure 4.** The plots of single-particle excitation spectrum given by Eq.(46) versus dimensionless wave vector for chemical potential µ = 0 (unbroken horizontal straight line). The other parameters are b = 0.95, $t_{d1}$= 1, $\varepsilon_f$ = -0.02, V = 0.10; $V_p = \left(\frac{Vab}{d}\right) = 0.04$. The hopping integral (a) $t_{f1} = 0.31$ (metallic bulk) and (b) $t_{f1} = -0.31$ (insulating bulk). The Rashba coupling strength (a) $\lambda_R = 0.80$ and (a) $\lambda_R = 0.40$. Since the conduction bands are partially empty, the surface state will be metallic in both (a) and (b).

We have plotted the surface state single-particle excitation spectrum (46) as a function of the dimensionless wave vector( *aK*) in Figure 4 for the considerably high Rashba coupling ($\frac{\lambda_R}{t_{d1}}$~0.5).The influences of the coupling on the spectrum is quite conspicuous if one compares this figure with figure 3. The 'gaplessness at wave vector K = 0' is the striking feature of the spectrum in 4(a). In contrast, for the insulating bulk in 4(b), this feature is also manifested at K = ±1. The wider spin-splitting under the influence of the Rashba interaction are basically responsible for prominent gapless Dirac spectrum including the cones. The TKI surface, therefore, comprises of 'helical liquids'**[38]** in the presence of the Rashba impurities. The access to the gapless Dirac spectrum leads to spin-plasmons with the usual wave vector dependence $q^{1/2}$. As we have noted above, the Rashba coupling does not impair the Kondo screening and does not affect the QCP for the bulk.

## 4. DISCUSSION AND CONCLUDING REMARKS

In the present communication we have started with PAM- a model for a generic TKI. The model itself and its extensions are still relevant for the theoretical condensed-matter physics. The examples of extension are those including the on-site interaction between *d*- and *f*-electrons and the nearest-neighbour interaction between *f*-electrons. A stronger version of this on-site interaction is found to destroy the Kondo state and narrow the intermediate valence regime **[5]**, while the latter allows to affect the stability of the magnetic ground state in the Kondo regime **[6]**. We also have introduced an extension to the model here by way of involving the magnetic exchange interaction. The interaction is introduced in the most direct route using only the spin degrees of freedom. We have demonstrated that the exchange field can be used to open a gap at the surface state dispersion. The field, in fact, is expected to play a bigger role, such as in the efficient tuning of the bulk band gap, the plasmon frequency, and so on. Looking backward, we observe there are many unsettled issues. For example, the problem of hybridization of plasmons with optical surface phonons that is likely to occur when TKI is surrounded by a material other than air has not been addressed. We need to suggest the ways and means to curb such a loss.

In conclusion, the inclusion of Rashba impurities in TKI surface gives rise to surface helical states where spin and momentum directions are locked to each other. In the proximity of a superconductor or a magnet, several interesting phenomena, such as the appearance of Majorana anyons, topological Faraday/Kerr effect, fractional charge for the quantum spin Hall effect in 2D, etc. may occur. These offer promising prospects for spintronic applications. Though immense progress has been made in this area over the past decade, never-the-less, we believe that our work may cast new light onto the investigations of how electron correlations and magnetic disorder influence the helical liquid. There are many challenges in the processing of these exotic materials to use the metallic/insulating states in functional devices, and they present great opportunities for the materials science research communities.

**Appendix A**

The scheme to calculate the averages $\langle d^{\dagger}_{k,\zeta} d_{k,\zeta}\rangle, \langle bs^{\dagger}_{k,\zeta} bs_{k,\zeta}\rangle$, etc. have been shown in this appendix. For this purpose, we proceed with finite-temperature formalism. Since the Hamiltonian is completely diagonal one can write down easily the equations for the operators $\{d_{k,\zeta}(\tau), s_{k,\zeta}(\tau)\}$, where the time evolution an operator $O$ is given by $O(\tau)=\exp(\aleph\tau)\, O\, \exp(-\aleph\tau)$, to ensure that the thermal averages in the equations above are determined in a self-consistent manner. The Green's functions $G_{sb}(k\zeta, k\zeta,\tau) = -\langle T_{\tau}\{d_{k,\zeta}(\tau)d^{\dagger}_{k,\zeta}(0)\}\rangle$, $F_{sb}(k\zeta, k\zeta,\tau) = -b^2\langle T_{\tau}\{s_{k,\zeta}(\tau)d^{\dagger}_{k,\zeta}(0)\}\rangle$, etc., where $T_{\tau}$ is the time-ordering operator which arranges other operators from right to left in the ascending order of imaginary time $\tau$, are of primary interest. We find

$$G_{sb}(k\uparrow, k\uparrow, \tau \to 0^+) = u^{(-)2}_{k,+}(e^{\beta(\epsilon^{(-)}_-(k)-\mu)}+1)^{-1} + u^{(-)2}_{k,-}(e^{\beta(\epsilon^{(-)}_+(k)-\mu)}+1)^{-1}, \quad (A1)$$

$$G_{sb}(k\downarrow, k\downarrow, \tau \to 0^+) = u^{(+)2}_{k,+}(e^{\beta(\epsilon^{(+)}_-(k)-\mu)}+1)^{-1} + u^{(+)2}_{k,-}(e^{\beta(\epsilon^{(+)}_+(k)-\mu)}+1)^{-1}, \quad (A2)$$

$$F_{sb}(k\uparrow, k\uparrow, \tau \to 0^+) = (u^{(+)2}_{k,+} - v^{(+)2}_k)(e^{\beta(\epsilon^{(+)}_-(k)-\mu)}+1)^{-1} + (u^{(+)2}_{k,-} + v^{(-)2}_k)(e^{\beta(\epsilon^{(+)}_+(k)-\mu)}+1)^{-1}$$

$$+ v^{(+)2}_k(e^{\beta(\epsilon^{(-)}_+(k)-\mu)}+1)^{-1} - v^{(-)2}_k(e^{\beta(\epsilon^{(-)}_-(k)-\mu)}+1)^{-1}, \quad (A3)$$

$$F_{sb}(k\downarrow, k\downarrow, \tau \to 0^+) = (u^{(+)2}_{k,-} + v^{(+)2}_k)(e^{\beta(\epsilon^{(-)}_+(k)-\mu)}+1)^{-1} + (u^{(+)2}_{k,+} - v^{(-)2}_k)(e^{\beta(\epsilon^{(-)}_-(k)-\mu)}+1)^{-1}$$

$$- v^{(+)2}_k(e^{\beta(\epsilon^{(+)}_-(k)-\mu)}+1)^{-1} + v^{(-)2}_k(e^{\beta(\epsilon^{(+)}_+(k)-\mu)}+1)^{-1}. \quad (A4)$$

,
For the averages $\langle d^{\dagger}_{k\uparrow} bs_{k,\downarrow}\rangle$ and $\langle bs^{\dagger}_{k,\uparrow} d_{k\downarrow}\rangle$, respectively, we obtain $\frac{V(s_x + i s_y)}{\varepsilon_-(k,b,\lambda,\xi)}\left[(e^{\beta(\epsilon^{(-)}_-(k)-\mu)}+1)^{-1} - (e^{\beta(\epsilon^{(-)}_+(k)-\mu)}+1)^{-1}\right]$ and $\frac{V(s_x + i s_y)}{\varepsilon_+(k,b,\lambda,\xi)}\left[(e^{\beta(\epsilon^{(+)}_-(k)-\mu)}+1)^{-1} - (e^{\beta(\epsilon^{(+)}_+(k)-\mu)}+1)^{-1}\right]$. These averages involving hybridization parameter $V$ are ultimate signature of the Kondo insulating state, where there is precisely one conduction electron paired with an impurity spin. In the zero-temperature limit the Fermi functions $(e^{\beta(\epsilon(k)-\mu)} + 1)^{-1}$ will be replaced by the Heaviside step function $\theta(\mu - \epsilon(k))$. Here

$$u_{k,\pm}^{(\zeta)2} = \frac{1}{2}[1 \pm \frac{(2\xi + \epsilon_k^d - b^2\epsilon_k^f + \lambda)}{2\{\sqrt{\frac{(2\xi + \epsilon_k^d - b^2\epsilon_k^f + \lambda)^2}{4} + 4V^2b^2(s_x^2 + s_y^2 + s_z^2)}\}}], \tag{A5}$$

$$v_k^{(\sigma)2} = \frac{-2V^2b^2s_z^2}{\sqrt{\frac{(2\xi + \epsilon_k^d - b^2\epsilon_k^f + \lambda)^2}{4} + 4V^2b^2(s_x^2 + s_y^2 + s_z^2)}\left[\frac{(2\xi + \epsilon_k^d - b^2\epsilon_k^f + \lambda)}{2} + \sigma\sqrt{\frac{(2\xi + \epsilon_k^d - b^2\epsilon_k^f + \lambda)^2}{4} + 4V^2b^2(s_x^2 + s_y^2 + s_z^2)}\right]}, \tag{A6}$$

$$\varepsilon_-(k,b,\lambda,\xi) = \sqrt{\frac{(2\xi + \epsilon_k^d - b^2\epsilon_k^f + \lambda)^2}{4} + 4V^2b^2(s_x^2 + s_y^2 + s_z^2)}, \tag{A7}$$

$$\varepsilon_+(k,b,\lambda,\xi) = \sqrt{\frac{(2\xi + \epsilon_k^d - b^2\epsilon_k^f + \lambda)^2}{4} + 4V^2b^2(s_x^2 + s_y^2 + s_z^2)}, \tag{A8}$$

$$\epsilon_\alpha^{(\zeta)}(k) = -\frac{(\epsilon_k^d + b^2\epsilon_k^f - \lambda)}{2} + \alpha\sqrt{\frac{(2\xi + \epsilon_k^d - b^2\epsilon_k^f + \lambda)^2}{4} + 4V^2b^2(s_x^2 + s_y^2 + s_z^2)}, \tag{A9}$$

$$\epsilon_k^d = [2t_{d1}\, c_1(k) + 4t_{d2}\, c_2(k) + 8t_{d3}\, c_3(k)], \tag{A10}$$

$$\epsilon_k^f = [-\epsilon_f + 2t_{f1}\, c_1(k) + 4t_{f2}\, c_2(k) + 8t_{f3}\, c_3(k)], \tag{A11}$$

and $\alpha = 1\,(-1)$ for upper band (lower band), $\zeta = \pm 1$ labels the eigenstates ($\uparrow, \downarrow$) of $\zeta_z$.